\documentclass[conference]{IEEEtran}
\usepackage{cite}
\usepackage{amsmath,amssymb,amsfonts}
\usepackage{algorithmic}
\usepackage{graphicx}
\usepackage{textcomp}
\usepackage{xcolor}
\usepackage{url}
\usepackage{booktabs}
\usepackage[caption=false,font=footnotesize]{subfig}
\usepackage{soul}

\begin{document}


\newcommand{\tokenName}[1]{{\small\texttt{#1}}}
\newcommand{\tokenAmount}[2]{{#1\,\tokenName{#2}}}

\newcommand{\hTag}[1]{{\small\texttt{\##1}}}
\newcommand{\cTag}[1]{{\small\texttt{\$#1}}}

\newcommand{\cAddr}[1]{{\small\texttt{#1}}}

\newcommand{\madval}[1]{\mbox{\textsc{mad}\,=\,#1}}
\newcommand{\medval}[1]{\mbox{$\tilde{x}$\,=\,#1}}
\newcommand{\medmadval}[2]{\medval{#1}; \madval{#2}}

\newcommand{\multinomchi}[4]{$\chi^{2}_{#1}$\,=\,#2; $p$\,#3\,#4}

\title{Toward Blockchain-based Fashion Wearables in the Metaverse: the Case of Decentraland}

\author{
\IEEEauthorblockN{Amaury Trujillo}
\IEEEauthorblockA{\textit{IIT-CNR} \\
Pisa, Italy \\
amaury.trujillo@iit.cnr.it}
\and
\IEEEauthorblockN{Clara Bacciu}
\IEEEauthorblockA{\textit{IIT-CNR} \\
Pisa, Italy \\
clara.bacciu@iit.cnr.it}
}

\maketitle

\begin{abstract}

Among the earliest projects to combine the Metaverse and non-fungible tokens (NFTs) we find Decentraland, a blockchain-based virtual world that touts itself as the first to be owned by its users.
In particular, the platform's virtual wearables (which allow avatar appearance customization) have attracted much attention from users, content creators, and the fashion industry.
In this work, we present the first study to quantitatively characterize Decentraland's wearables, their publication, minting, and sales on the platform's marketplace. 
Our results indicate that wearables are mostly given away to promote and increase engagement on other cryptoasset or Metaverse projects, and only a small fraction is sold on the platform's marketplace, where the price is mainly driven by the preset wearable's rarity. Hence, platforms that offer virtual wearable NFTs should pay particular attention to the economics around this kind of assets beyond their mere sale.

\end{abstract}

\begin{IEEEkeywords}
metaverse, virtual fashion, non-fungible tokens
\end{IEEEkeywords}

\section{Introduction}
\label{sec:introduction}

The idea of the Metaverse as a fully immersive future iteration of the Internet has fueled the imagination of many tech enthusiasts for a few decades now, with the latest advancements in extended reality (XR) having further ignited public interest.
The fashion industry is no exception; in recent years there has been an increasing participation in Metaverse-related initiatives, ranging from independent designers to famous commodity and luxury brands~\cite{gonzalez2022digital}. After all, avatar customization is one of the main enjoyments of users and one of the main factors of purchase intent in virtual worlds~\cite{bleize2019factors}, with the economic potential being highly enticing to the fashion industry~\cite{gonzalez2022digital}.
Also recently, non-fungible tokens (NFTs) ---units of cryptographic data that represent a unique asset via blockchain technology--- are being used to certify ownership of digital objects such as virtual garments.

Arguably, the first and most influential Metaverse-related project that utilized NFTs is Decentraland, a virtual world based on the Ethereum blockchain in which users can create, experience, and monetize their content and assets (including wearables for avatars) via the in-platform cryptocurrency, called \tokenName{MANA}.
The platform is managed through a Decentralized Autonomous Organization (DAO) ---a blockchain-based system that enables self coordination and governance~\cite{hassan2021decentralized}---
based on \tokenName{MANA} ownership.
In addition, users can also ``buy'' several kinds of NFTs, namely parcels of land; groups of parcels called estates; unique names; as well as wearables and the recently introduced emotes, items to customize the appearance and animation sequences of avatars, respectively.

Decentraland started in 2015 and launched in February 2020 to a moderate success. In 2021, there was a surge in the public interest in NFTs caused by several high-profile sales~\cite{trujillo2022surge}, giving rise to a sudden increase in the value of \tokenName{MANA}. A few months later, ``Facebook Inc.'' rebranded itself  into ``Meta Platforms'', drawing even more attention to the Metaverse in general, and to Decentraland in particular.
In view of this newfound popularity, Decentraland was host to the first Metaverse Fashion Week (MVFW) in March 2022, which involved a wide variety of brands and several in-world fashion events. However, the 2022 cryptowinter (i.e., a remarkable devaluation of cryptoassets) greatly affected Decentraland.
Such a volatility, together with many uncertainties surrounding NFTs, brought into discussion the approach used by Decentraland for its economy and the ownership of the in-platform digital assets, particularly wearables, with potential ramifications for other Metaverse-related projects.

Herein, we thus characterize the publication, minting, and marketplace sales of Decentraland's wearables.
Our work's main contribution is ---to the best of our knowledge--- the first quantitative characterization in literature of blockchain-based digital wearable assets on a virtual world.

\section{Background and Related Work}
\label{sec:background}

By Metaverse, we refer to a shared \textit{universe} of interoperable virtual worlds, with these in turn being ``[s]hared, simulated spaces which are inhabited and shaped by their inhabitants who are represented as avatars''~\cite{girvan2018virtual}.
We are still in an early development phase of the Metaverse, but a plethora of independent virtual worlds have already been created in the last decades, such as Second Life, or massively multiplayer online (MMO) games such as World of Warcraft, Minecraft and Fortnite.
Nevertheless, these have been subject to much criticism in terms of the governance and \emph{ownership} of in-world assets: a new form of virtual feudalism in Second Life~\cite{grimmelmann2009virtual}, conflicts of agency and ownership in World of Warcraft between publisher and gamers~\cite{glas2010games}, and an overall separation of content and platform ownership in virtual worlds to the detriment of users~\cite{zhou2018ownership}.
As a response, a wave of blockchain virtual experiences with decentralized governance and assets emerged, of which Decentraland is a prominent example.

Still, most previous research on Decentraland either focus on landholding assets~\cite{dowling2022fertile, goldberg2021land}, or take the platform's tokens as a whole within a larger NFT ecosystem~\cite{nadini2021mapping}. As far as we know, Decentraland's wearables have not been yet studied in literature.
Fashion within virtual worlds, on the other hand, has been studied for several years now, mainly on the still on-going Second Life (released in 2003), which popularized the idea of customizable virtual worlds and avatars~\cite{bardzell2010virtual}, but who many argue failed to meet the expectations of the general public~\cite{rodriguez2022dressing}. Nonetheless, within Second Life there is still a strong economy based on avatar customization and many users have found a new passion (and in some cases a source of income) by becoming in-world ``fashion designers'' of avatar clothing and accessories\cite{coellstorff2015}. 
In the last few years, however, the fashion industry has begun to focus more on highly popular MMO games ---mainly through branded \emph{skins}, i.e., in-game character cosmetic options--- with a growing interest on more recent blockchain-based virtual worlds~\cite{rodriguez2022dressing}. For instance, many important luxury brands have started experimenting with NFT collectibles~\cite{joy2022digital}, e.g., participating in initiatives such as the MVFW or also creating Decentraland weareables~\cite{gonzalez2022digital}.

\begin{figure}[!t]
\centering
\includegraphics[width=\columnwidth]{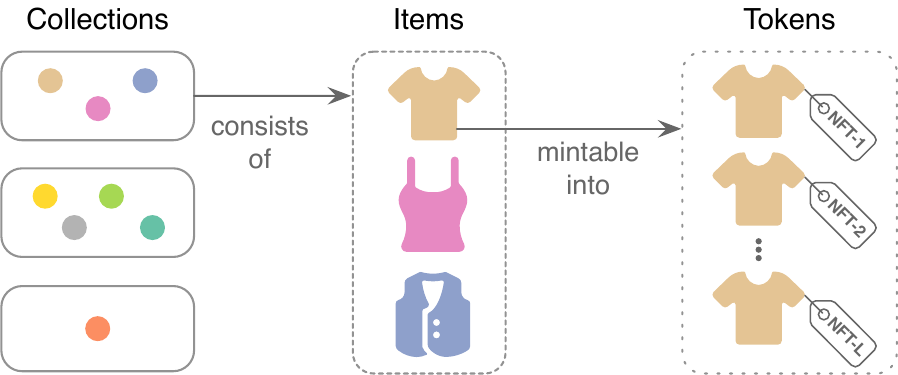}
\caption{After a Decentraland wearable collection is approved by a curation committee, each distinct item in it becomes available for \emph{minting} into NFTs up to the item's rarity limit chosen by the collection creator.}
\label{fig:collections}
\end{figure}

These wearables consist of items in the form of clothing, accessories, and partial or complete body features that change the appearance of avatars. Each wearable item belongs to a collection and can be \emph{minted} into unique NFTs, i.e. items themselves cannot be bought or sold, only the tokens minted from them (see Figure~\ref{fig:collections}).
First, collection creators must pay a fee so that it can be vetted by a curation committee (elected by the project's DAO).
If approved, the collection items are available for minting. Afterward, creators can either mint tokens and transfer them to any account in the blockchain network, or put their items on sale for minting on the official marketplace. Other users can obtain wearable tokens in three main ways: 1) by participating in an in-world event (e.g., a fashion show) that \textit{airdrops} (gives away) tokens; 2) by claiming tokens airdropped by an external project  (e.g., to reward its stakeholders); and 3) by buying tokens within or without the official marketplace. 
Regardless of the minting mechanism, the wearable NFTs are property of the holder; hence, it is also possible to use other trading platforms for secondary sales, such as OpenSea, the largest NFT marketplace. Herein we mainly focus on primary sales on Decentraland's marketplace.

Currently, there are two wearable versions in Decentraland's marketplace: v1 was implemented on the Ethereum network, with the first collection being published in October 2019. However, many concerns emerged within the community around Ethereum's scalability and volatile fees to process transactions. Hence, v2 wearables were developed for the Polygon network, a blockchain platform and an Ethereum-compatible sidechain that allows a more efficient processing of transactions~\cite{besanccon2022blockchain}. 
The last v1 collection was published in March 2021; the first v2 collection was published in June 2021, with v1 being deprecated but remaining usable in-world and available on NFT marketplaces. Therefore, users must be aware of the network in which they carry out secondary sales.

\section{Methods}
\label{sec:methods}

We collected various wearable data, from first record available up to December 31 2022, via Decentraland's query endpoints on TheGraph ---an indexing protocol and service for querying blockchain networks. These endpoints are used across the different applications within Decentraland and their implementations are open source.\footnote{\url{https://github.com/decentraland}} In detail, we retrieved metadata regarding wearable collections (name, status, creator, blockchain network) and items (name, category, rarity, price), the 3D asset metadata for v2 wearables (number of triangles, meshes, textures), in addition to all v2 items' mints on Polygon, as well as the primary and secondary NFT sales (buyer, seller, kind, price) within Decentraland.
To compare the monetary value of wearable items and sales over time, we first converted the price in \tokenName{MANA} of a given item or sale to \tokenName{USD}, based on data from Yahoo! Finance, and we then adjusted for inflation on a monthly basis to the \tokenName{USD} value in December 2022, based on the Consumer Price Index (CPI) by the Bureau of Labor Statistics of the United States.
Henceforth, all monetary value will be expressed in CPI-adjusted \tokenName{USD}.
To present results, we use the median ($\tilde{x}$) and median absolute deviation (\textsc{mad}) to describe distributions, and multinomial $\chi^2$ tests to compare proportions of wearable kinds between groups.
We used the R statistical software and its ecosystem for all of the analyses.

\section{Published Wearables}

\begin{figure}[tbp]
\centering
\includegraphics[width=\columnwidth]{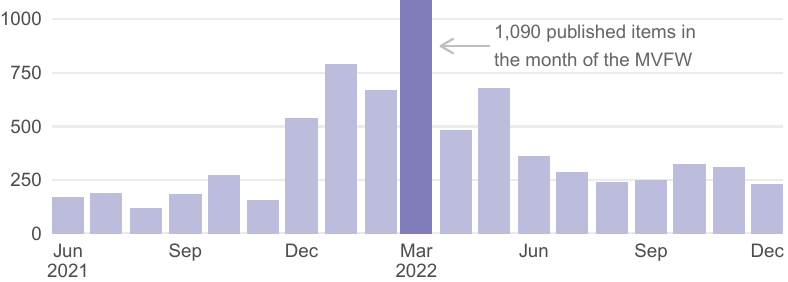}
\caption{Monthly publication count of approved v2 items. The month with the most publications corresponds to the Metaverse Fashion Week (MVFW).}
\label{fig:av2_monthly_items}
\end{figure}

For v1, there were only 41 collections (representing 436 items), whereas at the cutoff date there were 4,110 v2 collections (8,324 items) submitted, with most (88\%) being approved and published (3,756 collections with 7,323 items).
Concerning the 1,266 distinct v2 creator accounts, more than half (53.9\%) had only one collection; the most prolific creator by far had 79 collections with 616 items, mostly used as rewards by Decentral Games, a company whose main product is a poker game within Decentraland. All v1 collections consist of two items or more, with 19.5\% having more than ten; whereas the majority of v2 collections (68.8\%) consist of a single item.
The monthly publication of approved v2 items, (see Figure~\ref{fig:av2_monthly_items}) follows the rise and decline in \tokenName{MANA}, with the most prolific month being March 2022, around the MVFW.

The median asking price of v2 wearables at the cutoff date was \tokenAmount{30.64}{USD} (\madval{45.42}).
Interestingly, 987 items (13.4\% of the total published) have a zero price, but are not necessarily listed for sale, as these are usually used for airdrops or claimed as rewards.
Table~\ref{tab:wearable_rarity} presents an overview of v2 items by \emph{rarity}, which is set by their creators and limits the number of token supply that can be minted. The share of items (from the grand total) by rarity and version follows a bell-shaped distribution, with \textit{legendary} being the most popular in both versions. The relative minted supply (i.e. the proportion of tokens already minted) by rarity has instead a linear-log correlation, with the lower the limit the more supply minted.

\begin{table}
    \centering
    \caption{Overview of published v2 wearable items by rarity}
    \label{tab:wearable_rarity}
    \begin{tabular}{lrrrr}
        \toprule
        Rarity & Limit & Item share & Median price & Minted supply \\
        \midrule
            unique &      1  &  4.2\% & \$2003.04 & 0.879 \\
            mythic &     10  &  7.6\% &  \$156.70 & 0.683 \\
         legendary &    100  & 32.3\% &   \$41.46 & 0.452 \\
              epic &   1000  & 31.5\% &   \$28.20 & 0.224 \\
              rare &   5000  & 11.7\% &   \$11.69 & 0.179 \\
          uncommon &  10000  &  7.4\% &    \$7.30 & 0.157 \\
            common & 100000  &  5.3\% &    \$3.03 & 0.036 \\
        \bottomrule
    \end{tabular}
\end{table}

\begin{figure}[tbp]
\centering
\includegraphics[width=\columnwidth]{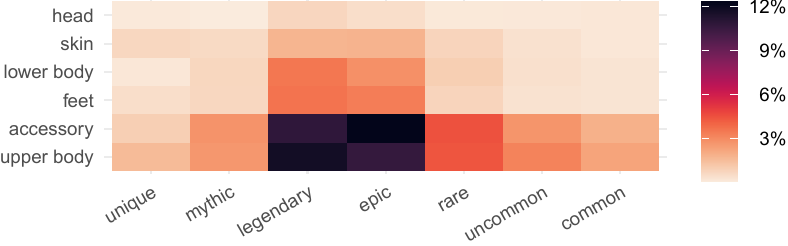}
\caption{Item share of published v2 wearables by \textit{category} and \textit{rarity}.}
\label{fig:item_share_by_category_and_rarity}
\end{figure}

Wearable 3D models can be adapted to different body shapes: female, male, or both (unisex). Between v1 and v2 there was a significant change in the share of items by body shape (\multinomchi{2}{86.9}{$<$}{.001}), with \textit{unisex} increasing by 12.7 percent points to 92.3\%, whereas \textit{female} and \textit{male} decreased to a respective share of 4.1\% and 3.6\%.
Other 3D model attributes refer to resources for its rendering, such as the number of triangles and  textures, which –––according to the guidelines--- should be limited by wearable \emph{category}. These are: upper body, lower body, feet,  accessory (i.e., earring, hat, helmet, etc.), head (i.e., facial hair, eyebrows, etc.), and skin (which changes the appearance of the whole avatar). For instance, no more than 5k triangles and 5 textures for \textit{skin}; no more than 1.5k triangles for all other wearables except \textit{accessory} (500 triangles) with at most 2 textures.
However, we found that these values are exceeded in 41\% of the items for triangles and 15\% for textures.
By far the most occurring item categories are \textit{upper body} and \textit{accessory}, with 36.6\% and 36\%, respectively. Based on Freeman's theta measure of association for ordinal and multinomial variables, \textit{rarity} and \textit{category} are very weakly associated ($\theta$\,=\,0.06).
Still, and as can be seen in the item share by these two variables in Figure~\ref{fig:item_share_by_category_and_rarity}, \textit{skin} has proportionally more \textit{unique} items (11\%) compared to the mean of the other categories (4.3\%).

\section{Token Mints}

The maximum token supply of v1 wearables was 246.2k, of which 70.9k tokens (28.8\%) have been minted. However, given the deprecation of v1, their minting has practically halted.
The maximum v2 token supply was 50.87M, of which only 3.56M tokens (7\%) have been minted. There were 1,225 v2 creators (96\% of the total) that published 6,9k items (94.3\%) minted at least once, or, stated differently, 50 creators (4\%) had 423 items (5.7\%) never minted as of the cut-off date.
Around 66.6\% of v2 creators with mints were involved on a primary sale on the marketplace, in which 8,953 buyers minted the tokens. Of the mints not associated with a primary sale, 682.3k mints (18.9\% of the grand total) were transfers from the creators to themselves, and the rest (80.8\% of the grand total) were transfers to other beneficiaries. Overall, there were 564.3k distinct beneficiaries of 3.56M minted tokens, with 354.7k beneficiaries (62.8\%) having received only one token, and 35.85k (6.4\%) ten or more.

\section{Marketplace Sales}

\begin{figure}[tbp]
\centering
\includegraphics[width=\columnwidth]{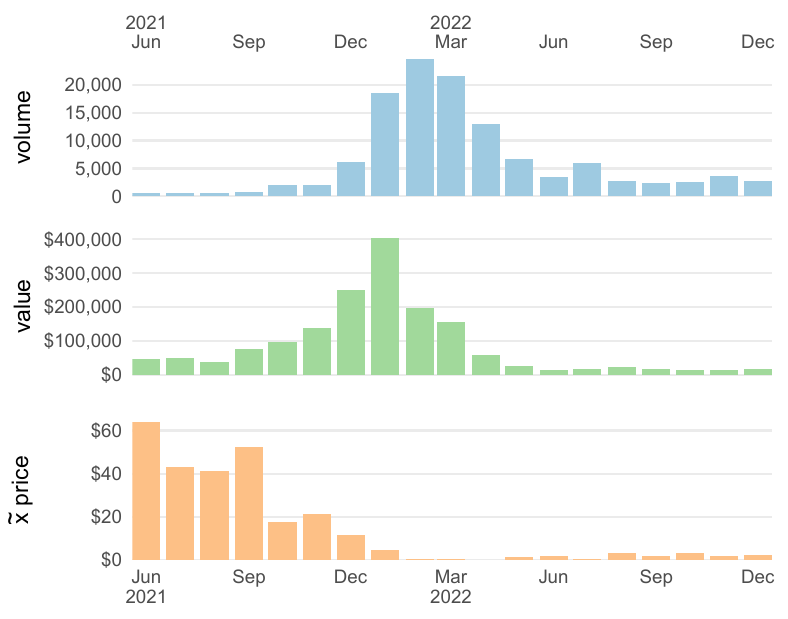}
\caption{Monthly v2 primary sales by volume, value, and median price.}
\label{fig:av2_primary_sales}
\end{figure}

Concerning the overall sales within the marketplace, v1 had 9.2K sales with a median sale price of \tokenAmount{3.1}{USD}, and v2 had 166.9K sales with a median sale of \tokenAmount{2.2}{USD}, with only 341 v1 sales being made since the introduction of v2. 
The number of v2 primary sales was 120.7k (72.4\% of the total), representing just 3.4\% of v2 mints; the median primary sale value was \tokenAmount{1.56}{USD} (\madval{2.32}). Despite both sales volume and value increasing at the end of 2021 and reaching their peak in early 2022, the median value of primary sales decreased remarkably during the second half of 2021 (see Figure~\ref{fig:av2_primary_sales}), most likely due to the highly increased supply of tokens. As for secondary sales, 9.7\% were bids (offers made for a token not listed for sale) and the rest were listings (of tokens publicly listed for sale), with an overall median value of \tokenAmount{3.03}{USD} (\madval{3.91}).

\begin{figure}[tbp]
\centering
\includegraphics[width=\columnwidth]{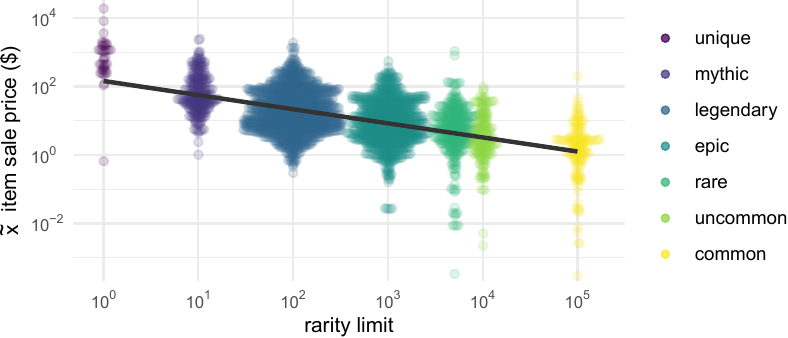}
\caption{Beeswarm plots of the median item sale price by rarity for primary sales (3.3k items). The black line depicts a simple elasticity linear model.}
\label{fig:primary_sales_price_and_rarity}
\end{figure}

The number of items with at least one primary sale was 3,332 (45.5\% of the total). We modeled the item median price of these sales by first selecting features intrinsic to the item (i.e., rarity, category, single or multiple item collection, 3D model complexity), and then using a forward stepwise regression linear model based on the Akaike information criterion (AIC). Table~\ref{tab:lm_primary_sales_price} reports the predictor coefficients of the final model, with $R^{2}_{adj}$\,=\,0.29, F\mbox{(4, 3327)}\,=\,341.3, $p$\textless.001.
Of these predictors, and based on a simple elasticity model, the most important is the item's \textit{rarity limit} ($R^2_{adj}$\,=\,0.24). As expected, the rarer the item, the more expensive it is. However, this predictor alone systematically underestimates the price of \textit{unique} items, as can be seen in Figure~\ref{fig:primary_sales_price_and_rarity}, hence the dummy predictor for this rarity in the final linear model.
Based on this model, the elasticity of the price with respect to the rarity limit is –0.389, and being \textit{unique} is associated with a +56.9\% price change. On the contrary, belonging to a single-item collection is associated with a –32.5\% change. Interestingly, the number of textures (and not triangles) is the most relevant 3D model metric, with an associated +5.7\% price change by texture.

\begin{table}
    \centering
    \caption{Linear model of the item median price for primary sales}
    \label{tab:lm_primary_sales_price}
    \begin{tabular}{lrrc}
        \toprule
        & \multicolumn{3}{c}{$\mathbf{log_{10}(\textbf{}{price})}$}\\
        \cmidrule(lr){2-4}
        Coefficient & Estimate & Std. Error & $p$\\
        \midrule
                       (Intercept) &  2.137 & 0.049 & \textless.001\\
          $log_{10}$(rarity limit) & –0.389 & 0.012 & \textless.001\\
         is single-item collection & –0.325 & 0.024 & \textless.001\\
                         is unique &  0.569 & 0.119 & \textless.001\\
                          textures &  0.057 & 0.013 & \textless.001\\
        \bottomrule
    \end{tabular}
\end{table}

In regards to secondary sales, there were 3,269 distinct items, of which 1,508 (46\%) did not have any primary sale within the marketplace, i.e., the respective tokens were minted by other means and then sold on the marketplace. Of the tokens minted via a primary sale, only 3.4\% were the object of a secondary sale. The tokens with multiple sales were 7,375, with 88.3\% having just two sales. Of the 840 tokens with three of more sales, 73 (8.6\%) have at least one round trip sale, i.e., an account eventually re-buys the token it previously sold. This is indicative of trade washing, a manipulative practice that affects most NFT markets in varying degrees~\cite{von2022nft}. The most blatant instance concerns the most sold token, with only two accounts trading it 15 times in just 18 days.

\section{Discussion}

\subsection{Change and standardization}

The change from the short-lived v1 to v2 wearables was substantial in terms of publication, implementation, and economics.
The initial version was, in part, victim to the sudden increase in interest on NFTs and on Ethereum in particular, which put into the limelight the network's transaction fees volatility, scalability issues. Hence, such a speedy transition was inevitable, but it was not without hiccups. For instance, changes in rendering of materials entailed a cumbersome procedure of 3D model rejection and re-submission, which puts in doubt the long-term functioning of in-world assets.
These issues are also a manifestation of the current frenzy around blockchain and the Metaverse.
Both are nowadays frequently associated with the so-called Web3, an envisioned decentralized iteration of the Web based on blockchain, and as it happened with the Web, to achieve such a vision it is necessary to first agree on common standards on a global scale.

Recently, several companies (including Meta and Microsoft) have founded the Metaverse Standards Forum, a platform for collaboration between standards organizations and businesses to promote an interoperable and open Metaverse.
The blockchain-based community reacted with the Open Alliance for the Metaverse (OMA3), with the intended goal of ensuring that virtual land, digital assets, ideas, and services follow the principles of Web3.
Decentraland is among the founding members of OMA3, which has now joined the Metaverse Standards Forum to work on 3D asset interoperability, digital asset management, privacy, cybersecurity and identity.
To add to the confusion, the Open Metaverse Foundation (by the Linux Foundation) has been recently launched. At the moment of writing all of these organizations had existed for less than a year; it remains to be seen if their missions are indeed reached.

\subsection{Wearables and the monetization of promotion}

Surprisingly, only a mere share of mints (3.4\%) is done via primary sales on the platform marketplace, and the volume of secondary sales on it is much lower than we had expected. Moreover, at the moment of writing (early 2023), there were only 3.2k tokens of v2 wearables listed as ``buy now'' (i.e., listed at a set price by the owner) on OpenSea, the most common external marketplace for Decentraland assets. From the minting data alone, it seems that the majority of tokens are given away as part of airdrops or awards to entice potential or existing users. Upon manual verification, this is certainly the case with the most prolific creators such as Decentral Games.

\begin{figure}[tbp]
\centering
\includegraphics[width=\columnwidth]{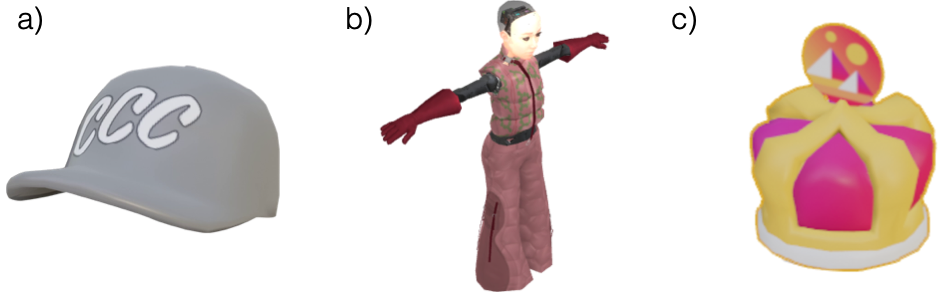}
\caption{Selection of outlier items: a) \textit{CCC - Hat - Gray}, the most minted with 99,961 mints; b) \textit{Sophia 42 - \#2}, the most complex 3D model with 8,657 triangles, 7 textures, and 6 meshes; c) \textit{The Royal Family Crown}, with the most primary sales being sold 7,781 times at zero price plus network fees.}
\label{fig:outlier_items}
\end{figure}

Further, for the most popular or complex items, it seems that the main scope is not earn money from the sales \textit{per se}, but to promote other products and services. For instance, Figure~\ref{fig:outlier_items} depicts: a) the top most minted item which was part of a give away to promote its creator; b) the item with the most complex 3D geometry used to promote Sophia The Robot; and c) the item with the highest number of primary sales, at zero price plus network fees.
Indeed, at the moment of writing MVFW 2023 hast just taken place, a testament of the in-world promotion appeal for the fashion industry.
Concerning desirability of wearables, based on our linear model the most important factor is the item's \textit{rarity}, especially if \textit{unique}, with other significant factors being the size of the collection and the number of textures of the 3D model.
It should be noted, however, that the term \emph{rarity} as used in Decentraland wearables differs considerably from the same term used on most NFT projects, in which it denotes the relative frequency of a token with a given set of characteristics. In the latter case, \emph{rarity} is also the most important factor for sale price, but with different and varying statistical distributions per project~\cite{mekacher2022heterogeneous}.

\subsection{Study limitations}
 
One of the main limitations of our work is the focus on Decentraland's marketplace: a creator that mints a token for themselves could put it for sale (within or without Decentraland's marketplace), transfer it to other accounts (e.g, as part of an airdrop or claimed awards), or simply keep it indefinitely. At the same time, Decentraland's marketplace is the entrance point and main setting for trading wearables on the platform, particularly for v2 primary sales, and it is the most common data source regarding Decentraland used in previous literature~\cite{nadini2021mapping, goldberg2021land, dowling2022fertile}.
Another limitation is the lack of data and analysis regarding the effective use of wearables within Decentraland's virtual world. For instance, it is plausible that many users own wearables that they have not worn in-world and perhaps will never do or will do for a very brief period of time. However, the on-chain data concerns only ownership of wearables, which is an interesting aspect by itself, but it is a faulty proxy of actual fashion as seen in-world.
In addition, despite the popularity and influence of Decentraland, it would be interesting to see if similar wearable dynamics are present on similar blockchain-based virtual worlds.

\section{Conclusion}

Decentraland's blockchain-based approach to virtual wearables has piqued the attention of digital fashion designers and consumers. Although items can be minted and sold within the platform's marketplace, our study shows that most are distributed by other means, mainly with the scope of promoting other projects and initiatives. For those sold within, the price is unsurprisingly primarily driven by their predefined rarity, with collection cardinality and number of textures also being important. However, being among the first in this domain has also brought problems for the platform regarding price volatility and long-term robustness of its technical implementations.
We thus hope that our work inspires others working on bringing the Metaverse into a vogue reality.

\bibliographystyle{IEEEtran}
\bibliography{references}

\end{document}